# Interoperable Architecture for Digital Identity Delegation for AI Agents with Blockchain Integration

David Ricardo Saavedra Martinez
Departamento de Ingenieria de Sistemas y Computación
Universidad de Los Andes - Bogotá D. C. Colombia
dr.saavedram1@uniandes.edu.co

*Abstract - This research proposes a conceptual and architectural framework for verifiable delegation in digital identity systems, designed to operate across centralized, federated, and self-sovereign identity (SSI) environments and to accommodate both human and AI agents. The core contributions are: (i) Delegation Grants (DGs) as first-class authorization artefacts, distinct from VCs, that encode bounded transfers of authority and enforce reduction of scope along delegation chains; (ii) the Canonical Verification Context (CVC), a protocol and format agnostic normalization model that represents any verification request as a single structured object; (iii) a layered architecture that separates trust anchoring, credential/proof validation, delegation and policy evaluation, and protocol routing via a Trust Gateway; and (iv) an explicit treatment of blockchain anchoring as an optional integrity layer, rather than as a structural dependency.*
.

## I. INTRODUCTION

Digital identity is the foundational mechanism through which entities, people, organizations, devices, and increasingly, autonomous agents, authenticate, authorize, and act across digital ecosystems. It defines who or what is interacting, what rights or obligations accompany that interaction, and how accountability and auditability are enforced afterward.

Despite decades of investment in public key infrastructures, federated identity management, and the more recent movement toward self-sovereign identity (SSI), there is still no universally adopted framework that guarantees interoperability, privacy, and accountability across domains. The global landscape remains a patchwork of incompatible systems, ranging from centralized authentication providers and federated protocols such as SAML and OpenID Connect, to decentralized technologies based on Decentralized Identifiers (DIDs) and Verifiable Credentials (VCs). These approaches coexist without a common semantic layer for trust, assurance, or delegation. Consequently, identity data is fragmented, interoperability depends on costly bridges, and users lose effective control over how and where their credentials are used or shared.

At the same time, digital identity is undergoing a second transformation with the emergence of autonomous software agents powered by artificial intelligence. These agents are expected to act on behalf of users, conducting transactions, accessing information, and executing policies, with the expectation of doing it without direct human supervision. Current identity infrastructures, built around explicit consent and manual credential presentation, were never designed for this paradigm. They lack a verifiable mechanism by which a human can delegate authority to an agent while preserving cryptographic sovereignty, defining clear limits of action, and maintaining traceability.

Without such a mechanism, trustworthy automation in critical domains such as e-government, finance, and healthcare remains infeasible. The absence of formal models for delegation and verification creates uncertainty about accountability, liability, and control in digital ecosystems increasingly driven by intelligent agents [1]. This thesis addresses precisely that gap: how to engineer a verifiable, interoperable, and auditable architecture of digital identity that supports secure delegation across heterogeneous and evolving technological environments. Some of them are:

### A. Centralized and Federated Models

The earliest digital identity systems relied on centralized models, where a single provider such as governmental, financial, or corporate issued and verified credentials. Although these systems are operationally efficient within their own domains, they introduce structural fragility and vendor dependence, as users trust and privacy hinge on a single authority.

Federated models, such as those based on SAML, OAuth 2.0, or OpenID Connect, improved usability through single sign-on and cross domain access. Yet, despite their practical success, they remain hierarchical and institutionally bound. In essence, federation distributes authentication, not sovereignty. Trust remains dependent on a limited number of identity providers, creating almost monopolies of digital trust and significant privacy implications.

### B. Decentralized and Blockchain Based Models

The introduction of blockchain and distributed-ledger technologies (DLTs) enabled a paradigm shift toward Decentralized Identity (DID) and Verifiable Credentials (VCs). Under the W3C's framework, trust emerges from cryptography and consensus rather than institutional mediation. [2]

In these systems, DIDs act as self controlled identifiers, while VCs encapsulate cryptographically signed claims that can be verified independently of the issuer. Together, they underpin Self Sovereign Identity (SSI), where users own and manage their digital credentials without relying on intermediaries.

However, SSI systems remain fragmented by design. Different DID methods (e.g., did:web, did:ion, did:ebsi) operate on distinct trust anchors and governance frameworks. Likewise, multiple credential formats like JSON-LD, JWT, SD-JWT use heterogeneous signature schemes, complicating selective disclosure and revocation. Thus, interoperability and standardization remain key obstacles to widespread deployment.

### C. Privacy and Zero-Knowledge Proofs

Recent progress in Zero Knowledge Proofs (ZKPs) [3] introduces the possibility of proving a statement's validity without exposing the underlying data. This represents a major advance in privacy preserving identity verification. For instance, a person can prove being over 18 without revealing their birth date, or an organization can prove a certification without sharing its internal documentation.

### D. Artificial Intelligence and Delegation

Existing frameworks offer limited delegation capabilities but remain centralized and session based. Their authorization tokens are temporary and related to a specific context, providing neither the fine grained scoping nor the cryptographic auditability required for distributed or AI driven environments. Consequently, these systems cannot formally express who delegated authority, to whom, and under what constraints, a gap that becomes increasingly relevant as autonomous systems scale.

Recent research aims to bridge this gap by extending familiar identity standards to support authenticated delegation [1] proposing the introduction of agent specific credentials and delegation tokens that explicitly link a human principal to an AI agent, defining the agent's scope of action and duration of authority. Similarly, work in the Self Sovereign Identity (SSI) community explores embedding delegation semantics within Verifiable Credentials (VCs), allowing independent verification of an agent's mandate without reliance on centralized providers [4]

### E. Regional Contrasts and Diversity of Ecosystems

Digital identity infrastructures have evolved differently across regions. Western models, built around pluralism and competition, face fragmentation among public and private actors. Each organization develops its own authentication systems, producing technical silos that must interoperate through complex agreements and middleware.

Conversely, China's digital ecosystem exemplifies a highly centralized model where major platforms, such as WeChat and Alipay [5], integrate identity, payments, and access control according with the government regulations. This creates an internally cohesive but tightly controlled environment, where verification and data sharing occurs seamlessly using, for example, mini apps managed by the same provider.

These contrasts reveal that digital identity is not a universal problem but a contextual reflection of how different societies balance innovation, control, and trust. The aim of this research is therefore not to replace existing paradigms, but to design a unifying, interoperable architecture that can bridge these divergent models and enable secure, auditable delegation of digital authority across them.

### F. Research approach

This research proposes a conceptual and architectural framework to enable verifiable delegation of digital identity across heterogeneous infrastructures. The approach integrates existing identity standards such as Decentralized Identifiers (DIDs), Verifiable Credentials (VCs), OpenID Connect (OIDC), and Zero Knowledge Proofs (ZKPs) into a cohesive model that supports both human driven and autonomous interactions.

The work is structured along two complementary dimensions:

1. Conceptual integration. A formal model is developed to describe how delegation relationships can be represented, verified, and revoked within identity systems. This includes defining the entities, trust boundaries, and invariants that govern authority transfer between humans, organizations, and intelligent agents.

2. Architectural design. The model is expressed through a modular reference architecture that bridges decentralized identity frameworks and web federated protocols. This architecture specifies how delegation tokens, verifiable credentials, and audit mechanisms interoperate across communication layers and trust domains.

By combining these elements, the thesis aims to define the missing interoperability layer for delegated digital identity, a layer that allows authority to be securely assigned, verified, and revoked across both centralized and decentralized systems. The result is an approach for trustworthy, auditable collaboration between humans and intelligent agents, supporting the evolution of digital ecosystems toward autonomy without compromising accountability or privacy.

## II. BACKGROUND-

In the context of the research, most identity systems can be abstracted into a three role trust model, sometimes referred to as the identity triangle: [6]

1. Issuer (Credential Provider): The entity that creates, signs, and issues verifiable credentials or assertions about a subject. [7]
2. Holder (or Subject): The entity to individual, organization, or agent that receives and controls credentials and can later present them as proofs. [8]
3. Verifier (Relying Party): The entity that validates the authenticity and integrity of the credential and uses it to make access or authorization decisions. [9]

This triad is a concept and invariant across identity standards such as OpenID Connect (OIDC), SAML, and W3C Verifiable Credentials (VCs). What differs among systems is how trust is anchored (institutional versus cryptographic), how credentials are transported and verified (web protocols versus distributed ledgers), and who controls the private keys and verification metadata.

### Identifiers, Attributes, and Credentials

At its core, a digital identity is a mapping between identifiers and attributes secured by cryptographic credentials:

• Identifiers uniquely reference an entity within a namespace. They may be centralized (e.g., email, national ID number), federated (e.g., sub claims in OIDC), or decentralized (e.g., DIDs).
• Attributes are descriptive properties of the entity such as name, role, or device fingerprint bound to the identifier.
• Credentials are attestations that cryptographically link identifiers to attributes, signed by an issuer using its private key.

This relationship can be expressed as a tuple (Identifier, Attribute, CredentialSignature) [10], where verification requires validating the issuer's signature with its public key. The resulting assurance provides integrity if the data has not been altered, authenticity (it originates from a trusted issuer), and non repudiation (the issuer cannot plausibly deny the issuance). This structure still being valid in diversed technical ecosystem such as the mentioned before in the specific case of China [11].

In decentralized environments, identifiers like DIDs resolve to DID Documents, which specify verification methods (e.g., Ed25519 or Secp256k1 keys) and service endpoints. This design decouples identity from any central registry while maintaining verifiability through shared cryptographic references. [12]

### Trust Establishment and Cryptographic Assurance

In regards Trust in digital identity systems, is important to mention that it is not absolute but compositional, meaning that it emerges from verifiable proofs at multiple layers, such as:

1. Proof of Possession: The holder must demonstrate control over the private key corresponding to the credential or identifier (e.g., via digital signatures or challenge response).
2. Proof of Integrity: The verifier checks that credential data has not been modified since issuance.
3. Proof of Origin: Metadata, issuer identifiers, or blockchain anchors confirm the credential's legitimate provenance.
4. Proof of Status: Revocation registries, status lists, or ledger transactions determine whether the credential remains valid.

Traditional Public Key Infrastructures (PKI) implement these proofs through hierarchical certificate authorities [13], while federated models like OIDC or SAML rely on HTTPS endpoints and JSON Web Signatures (JWS) to achieve similar guarantees. In decentralized systems, these assurances are distributed across ledger anchored key registries and cryptographic proofs of existence that remove the need for central trust intermediaries.

### Credential Lifecycle and Governance

Digital credential follows a lifecycle composed of four main stages [14]:

1. Issuance: The issuer generates and signs the credential, embedding metadata such as schema, expiration, and revocation endpoints.
2. Storage: The holder maintains the credential locally (wallets, secure enclaves) or within agent-mediated storage.
3. Presentation: The holder generates proof, often a derived, privacy-preserving version of the credential and transmits it to a verifier through a secure channel.
4. Revocation and Audit: The issuer or governing authority updates a registry or ledger to reflect credential status; verifiers check these registries before acceptance.

Governance frameworks define policies for credential schemas, verification rules, and trust-anchor management.

*Implications for Delegated Identity*

Up to this point, most of the developments related to Digital Identity shares a common approach, but the delegation to a third party, especially if it's an Artificial Agent introduces a fourth conceptual dimension: the delegator and delegate relationship, in which authority is transferred conditionally from one entity to another.

This extension requires reinterpreting traditional credential flows to include:

- explicit representation of delegation rights,
- cryptographic linkage between entities, and
- Lifecycle events for delegation activation, expiration, and revocation.

### III. PROPOSED FRAMEWORK.

This section lays out the intellectual and technical terrain of the research: it identifies the requirements for a verifiable, interoperable delegation mechanism that can function across heterogeneous digital identity infrastructures and that can accommodate both human actors and autonomous agents as first class participants in the trust fabric of the digital world.

The proposed model extends the well-known trust triangle of issuer, holder, verifier by into a Delegation tetrahedron, introducing a fourth role the Delegate as shown in Figure 1.

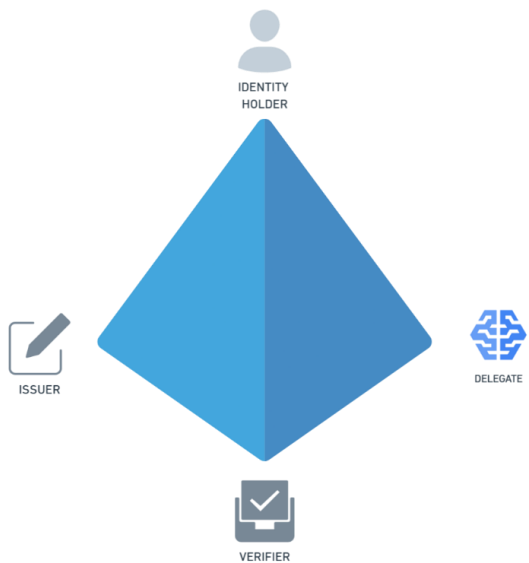

*Figure 1 Delegation tetrahedron*

In this expanded schema, a principal (human or organizational) may authorize another entity, whether a person, organization, or autonomous agent, to act on its behalf under a defined scope and within bounded parameters. This relationship is not implicit or session, as in OAuth or API key exchanges, but rather explicit, verifiable, and cryptographically anchored through a novel approach, that in the context of the research will be called Delegation Grant (DG). The DG becomes a first-class object in the identity ecosystem, capturing the semantics of authority transfer and ensuring that every delegated action can be verified as legitimate, traceable, and revocable.

Formally, the verification process can be abstracted as a function:

$$Verify(C, \pi, \Delta, P, S) \rightarrow \{true, false\}$$

Where:

- C: Set of Verifiable Credentials presented by the entity.
- $\pi$: Cryptographic proof demonstrating control over the credentials and identifiers.
- $\Delta$: An ordered chain of Delegation Grants [DG_1, ..., DG_n] linking the acting entity to a trusted root.
- P: Verification Policy defining trust anchors, allowed scopes, and assurance levels.
- S: Status context providing freshness guarantees for revocation checks.

The function returns true only if all credentials are valid and authentic, the proofs are correct, the delegation chain is well-formed and authorized for the action, and the entire state complies with P under status S

To operationalize this model, the architecture is structured around four logical layers, each serving a distinct but interdependent purpose:

*Layers or Sub models*

**L0. Trust Anchors and Resolution Layer**: Provides the cryptographic foundation by defining how identifiers, keys, and status registries are discovered and verified. This includes DID documents, JWKS endpoints, and ledger verification registries. It establishes the root of trust for all subsequent verification operations. There's also possible to provide additional anchoring for delegation chain using Blockchain for the Delegation Chain Registry.

**L1. Credential and Proof Layer:** Handles the ingestion and validation of verifiable credentials and their associated proofs, regardless of format. This layer supports VC-JWT, VC-LD, SD-JWT, and ZKP credentials, abstracting them into a common internal representation without altering signed content. It is here that credential authenticity, issuer legitimacy, and data integrity are verified.

**L2. Delegation and Policy Layer**: Introduces and enforces the Delegation Grant mechanism. This layer evaluates whether an action or credential presentation is authorized under a valid delegation chain, ensuring compliance with temporal, contextual, and hierarchical constraints. It also manages revocation, expiry, and scope limitations, forming the control logic that governs agent authority.

**L3. Trust Gateway and Interoperability Layer:** Functions as the dynamic routing and normalization engine for all verification requests. It detects the protocol or credential type whether OIDC4VP, VC-JWT, VC-LD, or SD-JWT and routes it to the appropriate verification handler. This layer ensures protocol neutrality and scalability by unifying verification results under a single canonical format known as the Canonical Verification Context (CVC).

*Canonical Verification Context*

The CVC serves as the core *of* the verification process. Regardless of the input format or transport protocol, all verification requests are normalized into this structure, which contains the credential metadata, proofs, issuers, status references, and delegation chain. This allows verifiers to apply consistent algorithms and policies without requiring specialized logic for each identity technology stack.

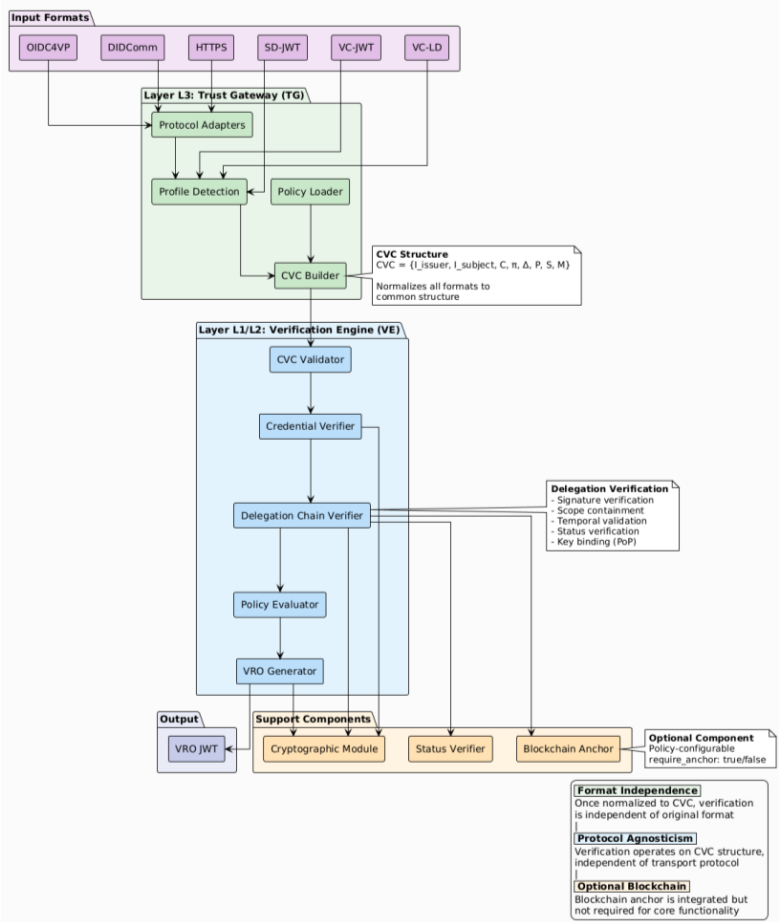

Figure 2 Theoretical Framework Diagram

The CVC is the normalized data structure exchanged internally between the architectural layers L0-L3. It serves as the canonical representation of a verification request, abstracting away all transport-specific and encoding differences (e.g., OIDC4VP vs VC-LD).

Formally, the CVC can be expressed as:

$$CVC = \{I_{issuer}, I_{subject}, C, \pi, \Delta, P, S, M\}$$

where:

- $I_{issuer}, I_{subject}$ are the resolved identifiers (DIDs, JWKS, or equivalent URIs).
- $C$ is the normalized credential set.
- $\pi$ represents normalized proof objects, including cryptographic type and verification material.
- $\Delta$ is the delegation chain (if present).
- $P$ is the applied verification policy.
- $S$ contains revocation and status references with timestamps.
- $M$ includes auxiliary metadata such as request identifiers, submission timestamps, and profile hints.

The CVC is instantiated by the Trust Gateway (L3) after protocol detection and before cryptographic verification begins. By converting all inbound data into this canonical form, the verification process in L1 (Credential/Proof Validation) and L2 (Delegation/Policy Evaluation) becomes protocol-agnostic and format-independent. This guarantees that the verification logic operates deterministically regardless of whether the input originated from an OAuth flow, a DIDComm exchange, a VC-LD presentation, or a ZK-proof system.

Together, these four layers form a modular, extensible, and testable conceptual framework that unifies web-federated and decentralized identity systems under a single verifiable delegation model.

*Trust Boundaries*

In a distributed identity ecosystem, trust is not continuous, it is segmented by boundaries where verification must be explicit and cryptographically enforced. Each interface between entities represents a potential point of failure if not properly secured. The architecture defines the following critical trust boundaries:

Holder-Issuer Boundary: The holder's trust in the issuer is established during credential issuance, but it is asymmetric: while the holder relies on the issuer's signature, the issuer has no guarantee of how the credential will later be used.

Holder-Delegate Boundary: The DG functions as a cryptographic contract binding the delegator's authorization to a delegate's verifiable identifier and key material. Both parties share accountability with the delegator by virtue of having signed the grant, and the delegate by virtue of having exercised it.

Delegate-Verifier Boundary: Privacy-preserving proof mechanisms constrain the information shared across this interface, preventing overexposure while retaining verifiability.

Verifier-Status Service Boundary: The verifier queries revocation and status information without revealing the credential subject or the verifier's identity..

Verifier-Trust Gateway Boundary: All verification traffic passes through this boundary. It is authenticated via mTLS or proof-of-possession tokens (DPoP) to ensure the authenticity of requests.

Audit Boundary: Only hashed summaries of verification events cross this line, allowing public accountability without the retention of personal data.

*Cryptographic Roles and Key Management*

Every entity must be able to prove identity, assert authorization, and validate provenance without relying on continuous connectivity or centralized mediation.

All entities are bound by verifiable key material, either through DID Documents, JWKS endpoints, or ledger-anchored proofs. Issuers and holders use signing keys to generate credentials and delegation grants. Delegates prove control over their corresponding keys to exercise granted rights. Verifiers, in turn, resolve public verification material and ensure that every signature, proof, and delegation is derived from valid key references.

- Key material is functionally specialized:
- Signing Keys create credentials, DGs, and proofs.
- Verification Keys are resolved by verifiers and status services.
- Revocation Keys maintain revocation registries and should be logically isolated to prevent linkage across transactions.
- Delegation Keys bind the delegator and delegate within the DG, ensuring non-transferability and controlled propagation of authority.

These operations are cryptographically sealed within the Status Context (S). Freshness guarantees, key rotations, and revocation events are continuously verifiable through ledger-anchored status entries or off-chain status endpoints.

By strictly separating cryptographic responsibilities across these roles, the system achieves verifiable independence because each participant can perform its function securely without requiring implicit trust in any other.

*Delegation Grant (DG)*

The DG follows the same trust semantics as Verifiable Credentials: the *issuer* (delegator) signs the object, and any verifier can check its validity using public verification material. However, unlike a credential, the DG's subject gains operational capability only when the DG is linked to a valid credential chain ($\Delta$) leading back to a trusted principal.

Based on the data structure, the scope enumerates verifiable permissions as *atomic, composable units* that can be reasoned over cryptographically and validated for subset containment along a delegation chain. During verification, each DG's scope is checked to ensure that for any child DG:

$$scope(DG_{i+1}) \subseteq scope(DG_i)$$

This guarantees monotonic reduction of authority with each delegation step, preventing privilege escalation.

In the same way, the key binding establishes a verifiable link between the DG and the cryptographic key that must be presented during authorization. Depending on the chosen method.

While the status object ensures revocability and freshness, aligning with W3C and SSI revocation mechanisms. Verification logic checks that the presented DG is unrevoked, cryptographically valid under its key_binding, and compliant with cache policies to ensure real time trust freshness without reintroducing centralization.

Additionally, to maintain compatibility across ecosystems, the model supports the three dominant proof formats in decentralized identity systems: JWS, SD-JWT, and Linked-Data (LD) Proofs. This flexibility allows the same verification logic to operate across web-native, SSI, and hybrid infrastructures.

*Validation and Chain Awareness*

To enhance the validation process of the DG is necessary to include a function of ordered sequences of authorizations that connect an initial delegator to a final delegate through verifiable, bounded transfers of authority. Because of it, it might be possible to verify the current state of the delegation. In response to this purpose, it's feasible to approach a blockchain based structure that will store the hash creation history without including all the DG information in order to protect the integrity of the information.

In that way, each link in this chain represents a cryptographically signed and policy-constrained relationship that must be independently valid while remaining consistent with the broader authorization context, independent of the technical approach to story or construct it.

This validation step does not simply confirm that a DG is well formed or correctly signed; it evaluates whether the entire chain of delegation is trustworthy, current, and compliant with the policies that define its scope. This ensures that authority flows securely through each step without escalation, duplication, or loss of accountability.

The whole validation can be described as follows

1. Structural Verification: Each DG must comply with the schema: required fields (issuer, subject, scope, proof, key_binding, and constraints) must be present and correctly formatted. Invalid or incomplete DGs are excluded from chain formation.
2. Cryptographic Verification: Each DG's proof must verify against the public key or verification method of its declared issuer. This guarantees the authenticity of every authorization step, regardless of the proof format used (JWS, SD-JWT, or LD-Proof).
3. Temporal and Contextual Coherence: The DGs in a chain must represent an active and consistent authorization path that is, each grant must still be valid at the time of verification, and no link may have expired, been revoked, or fall outside the defined policy scope.
4. Authority Continuity and Reduction: As delegation progresses, the scope of permission must remain equal to or narrower than that of the preceding DG. This enforces bounded delegation, ensuring that no intermediary can expand privileges beyond what was explicitly granted.
5. Revocation and Freshness: Each DG must resolve to an unrevoked state according to its defined status mechanism. Caching and validity periods must respect the freshness requirements specified by the verification policy.
6. Non-transferability and Key Binding. The acting delegate must demonstrate control of the key referenced in the DG's key_binding. This prevents DG reuse or substitution by unauthorized entities.
7. Chain Integrity: The complete delegation path must be acyclic (no repeated issuer-subject pairs) and must connect back to a trusted root or policy-defined authority. Verification concludes successfully only when all individual DGs form a coherent, validated path of trust.

*Trust Gateway (TG): Routing and Normalization Layer*

The Trust Gateway (TG) is the central interoperability mechanism of proposed architecture. It looks forward to enabling verification and delegation to operate seamlessly across heterogeneous identity ecosystems by abstracting underlying protocol differences and enforcing a unified verification process. Conceptually, the TG serves the same purpose that routers or gateways serve in network infrastructures: it ensures that requests, regardless of origin, encoding, or proof type are directed to the appropriate verification path, normalized into a common format, and evaluated consistently under a verifiable trust policy.

In digital identity terms, the TG acts as the control and normalization layer between diverse standards such as OAuth2, OpenID Connect (OIDC), OIDC4VCI/VP, Verifiable Credentials (VCs), and Delegation Grants (DGs). It provides a reliable mechanism to verify any credential or delegation artifact through a canonical verification model that remains protocol-agnostic, cryptographically sound, and auditable.

Functionally, the TG performs three essential tasks: i) Protocol Detection and Routing, ii) Normalization to the Canonical Verification Context (CVC), iii) Emission of Verifiable Results.

This design ensures that verification remains consistent, scalable, and future proof, regardless of the technologies used to issue, transport, or present identity artifacts, due to the fact the TG uses a dynamic adapter mechanism to identify and process incoming verification requests. Each adapter is a modular component responsible for parsing, validating, and transforming data according to its native profile. Detection and routing occur through declarative metadata inspection, not static configuration, ensuring long term extensibility.

Upon detection, the TG invokes the matching adapter, which performs Parsing meaning that deserialize and validate the native credential or DG object. Proof Verification where perform cryptographic verification specific to the detected proof suite and Normalization to map the result to the CVC structure for uniform evaluation downstream.

In this way the Trust Gateway represents the interoperability core of the verifiable delegation architecture. By instantiating every verification request as a Canonical Verification Context (CVC) before cryptographic validation, it eliminates dependency on specific identity protocols or proof formats. This architectural decision ensures that the verification logic in Layers L1 and L2 is consistent, scalable, and testable, regardless of input diversity.

*Verification Algorithms (Deterministic Core)*

The verification process constitutes the computational backbone of the proposed interoperable architecture. Once the Trust Gateway (L3) has normalized any incoming request into a Canonical Verification Context (CVC), the system performs verification deterministically, independent of the transport protocol (OIDC4VP, DIDComm, HTTPS) or credential encoding (VC-LD, VC-JWT, SD-JWT).

To ensure reproducibility and auditability, the verification function is defined as deterministic pseudocode rather than abstract mathematical notation. This formalization allows direct translation into executable reference code while preserving the architectural neutrality required for long term interoperability. The code follows a simple logic of validation that contains the following steps:

1. Structural Validation (Step 1): Guarantees canonical CVC integrity before any cryptographic operation.
2. Credential Verification (Step 2): Consolidates all proof mechanisms (JWS, LD-Proof, SD-JWT, ZKP) under a uniform interface. Each credential is checked for issuer authenticity, proof integrity, and temporal validity against its trust anchor in L0.
3. Presenter Binding (Step 3): Ensures the active subject (human or agent) controls the cryptographic key referenced in the credential or Delegation Grant. This step closes the possession gap and enforces accountability.
4. Delegation Chain Evaluation (Step 4): Validates bounded, acyclic propagation of authority along delegation chain. Optional blockchain anchoring as described earlier in Section 3.3 adds a tamper evident ordering mechanism without exposing DG contents.
5. Status and Freshness (Step 5): Synchronizes credential and DG revocation state, ensuring that cached data comply with freshness constraints defined by policy P.
6. Policy Enforcement (Step 6): Contextual rules such as assurance levels, trusted issuers, scope limits, and maximum chain depth are enforced deterministically.
7. Signed Verification Record (Step 7): Produces a portable, cryptographically signed Verification Result Object (VRO) serving as immutable evidence of verification outcome. This artifact integrates seamlessly with the rest of the architecture.

This design acknowledges the reality that digital identity standards will continue evolving, particularly as AI agents become more sophisticated in their delegation requirements. Together, these principles create a verification engine that is not merely theoretically sound but operationally viable and capable of enforcing consistent trust judgments today while adapting to the unknown identity challenges of tomorrow, thus fulfilling the research's scope of a durable, interoperable foundation for human-AI delegation in distributed digital ecosystems.

*Protocol Interoperability Profiles*

Through the CVC abstraction, any incoming verification request, regardless of origin or protocol, can be normalized into a standard structure, evaluated under the same logical flow, and produce identical verifiable outcomes. This section formalizes the interoperability profiles that define how this translation occurs in different ecosystems, demonstrating the architecture's ability to interoperate across federated web systems, decentralized identity frameworks, and hybrid configurations that combine both.

A. Web Federated Identity Environments

In web-federated contexts, trust relationships are typically mediated by OpenID Connect (OIDC) or OAuth 2.0 frameworks. Within this profile, the Trust Gateway functions as a validation proxy that converts the tokens into normalized CVC elements. It extracts the issuer (iss), subject (sub), cryptographic proof (JWS signature or SD-JWT disclosure chain), and temporal claims (iat, exp), mapping them to the fields expected by the verification core. Once canonicalized, the verification function treats the credential identically to any other verifiable artifact, applying the same proof validation, key resolution, and policy enforcement steps.

This approach preserves the operational semantics of web identity systems while introducing verifiability and delegation semantics that traditional OAuth infrastructures lack. It allows for the coexistence of human users, organizations, and autonomous agents under the same authorization structure.

B. Decentralized and Self-Sovereign Identity Frameworks

Here entities independently hold and present verifiable credentials under decentralized trust assumptions. Credentials are typically expressed in JSON-LD format, signed with Linked Data Proofs, and anchored in decentralized identifiers (DIDs) resolvable through blockchain or web registries.

The Trust Gateway accepts verifiable presentations (VPs) that include one or more credentials issued by autonomous or institutional entities. Upon reception, the gateway canonicalizes the JSON-LD graph into deterministic RDF datasets, verifies the cryptographic proofs (Ed25519 or BBS+), and maps the claims and proofs into the CVC's C and cryptography fields.

C. Hybrid Interoperability Environments

Corporations, governments, and autonomous agents must be able to use both web federated and decentralized infrastructures if business case requires it. Because of it, the Hybrid Profile bridges these environments by allowing mixed verification of federated credentials, decentralized proofs, and delegation artifacts under one canonical verification flow. In this configuration, Delegation Grants (DGs) are encoded as SD-JWTs. This format provides compactness, native JSON compatibility, and selective disclosure, enabling DGs to be validated alongside OIDC or SSI credentials without requiring additional protocol layers.

D. Profile Discovery and Policy Governance

To facilitate automation and governance, verifiers expose their operational capabilities through a machine readable descriptor, allowing clients and agents to discover supported profiles, proof types, and policy identifiers. This descriptor, served through a standardized endpoint defines which input formats, holder-binding mechanisms, and verification rules are accepted.

*Security and Privacy Model*

A fundamental criterion for any digital identity architecture is that it must balance verifiability, privacy, and control. [9] The security and privacy model outlined here builds directly upon the layered trust architecture (L0-L3) and the deterministic verification mechanisms introduced earlier. It aims to demonstrate that interoperability, achieved through protocol abstraction, does not come at the cost of reduced assurance, confidentiality, or accountability.

### Threat Model and Systemic Assumptions

The system operates under the assumption of potentially untrusted networks and participants. Adversaries may intercept, replay, or modify credential presentations, attempt to forge delegation chains, or exploit protocol downgrades. Internal actors such as compromised delegates or malicious verifiers may try to escalate privileges, correlate identities, or misuse cached data.

The framework mitigates these risks through a fail closed trust model, based on the principle where any unresolved or unverifiable element whether a missing key, expired proof, or stale status results in an explicit failure. All entities must be resolvable through the L0 trust anchor layer, which serves as the immutable basis for verifying DIDs, key material, and credential registries. The verification pipeline assumes authenticated transport channels, and relies solely on open, peer-reviewed cryptographic primitives. No single actor, including the verifier or gateway, can unilaterally alter verification outcomes than can be guaranteed through some decisions such as: i) Delegation Grants (DGs) can only narrow and never expand the authority granted by preceding links. Ii) The optional blockchain anchoring mechanism extends accountability beyond organizational boundaries. Iii) Determinism and Verifiability: Identical inputs always yield identical outputs, and the verifier never assumes correctness in the absence of proof.

### Privacy and Data Minimization

Verification requires only the cryptographic evidence necessary to confirm the validity of a claim, not the full disclosure of underlying attributes. Selective disclosure mechanisms, particularly SD-JWT and BBS+, allow entities to reveal the minimal subset of information required for policy compliance, ensuring privacy by design. The Trust Gateway processes such data ephemerally: once verification concludes, transient representations of the CVC are discarded or securely hashed. Importantly, when blockchain anchoring is used, only cryptographic commitments, not the actual credential contents, are published on-chain.

The framework also incorporates privacy protection through contextual isolation: verification sessions are not reusable, identifiers are pseudonymous where possible, and verifier logs contain only signed Verification Result Objects (VROs) devoid of personal information. This design aligns with the principle of proportional disclosure and satisfies contemporary regulatory frameworks such as the EU's GDPR and the emerging ISO/IEC 27560 standards for privacy-preserving identity systems.

### Threat Mitigation and Operational Safeguards

The system employs multiple strategies to mitigate common and advanced attack vectors:

- Replay and Impersonation Protection: Timestamp and nonce validation prevent reuse of old credentials or DGs. Key binding validation ensures that possession of data does not imply authorization to act.
- Scope and Privilege Containment: The verification algorithm enforces explicit scope reduction at each delegation hop, eliminating the possibility of uncontrolled privilege propagation.
- Correlation Resistance: Session isolation, ephemeral identifiers, and minimal persistent state prevent verifiers or external observers from linking multiple presentations to the same entity.
- Transparent Auditability: Each verification emits a signed VRO that records the decision context without leaking sensitive content.

### *Ethical and Governance Implications*

From an ethical perspective, the redistribution of authority enhances user autonomy and organizational accountability but also requires new governance models for policy definition, revocation management, and dispute resolution. The framework thus positions blockchain not merely as a storage mechanism but as a mechanism of institutional transparency, where authority is both decentralized and verifiable.

Nevertheless, it's important to mention that during the development of the research, more ethical and philosophical questions have arisen in order to understand the limit of digital identity to represent human nature, desire and representation. In that sense is important to acknowledge the lack of real representation from AI Agents due to its merely operational behave.

### *Limitations, Design Trade offs*

A first constraint arises from resolution latency at L0. DID document retrieval and JWKS lookups are network-bounded; while caching reduces average cost, freshness requirements in policy *P* cap cache lifetimes, and cold starts can dominate end-to-end latency. This is a deployment, not a model, limitation; nevertheless, it influences the observed performance envelope. A second constraint is standards drift. Profiles involving SD-JWT, OIDC4VCI/VP, and BBS+ are still maturing across implementations. The verifier mitigates drift by isolating these concerns behind adapters, but long-lived deployments must pin versions and track changes.

There is a deliberate balance between proof expressiveness and computational cost. LD-Proofs with BBS+ enable graph level selective disclosure and unlinkability, but they are heavier than compact JWS; JWTs and SD-JWTs are efficient and widely deployed but less expressive semantically. The profiles preserve both options and keep the verifier neutral by normalizing all artifacts into the CVC. Similarly, delegation integrity can be strengthened by optional anchoring of chain commitments; this improves tamper evidence and cross domain auditability at the cost of write operations and potential confirmation delays on public ledgers. The model treats anchoring as an integrity enhancement, not a dependency, so deployments can enable it where auditability outweighs cost.

In any case is important to point that even when the main drive to explore Blockchain integrations was due to the first interest of orchestrating the overall architecture through this technology, but during the development of the propose it was much more clear that even when it can bring advantages such as the mentioned previously, it might be more prudent to keep it as a possibility and not a required component.

Privacy is protected by selective disclosure and data minimization, yet residual risks remain. Status queries and anchor reads may leak access patterns if not batched or proxied; capability descriptors increase transparency but also reveal supported algorithms and policies. These concerns are addressed by short lived sessions, minimal persistent state, and VROs that record decisions without exposing subject attributes..

Determinism and fail closed semantics, central to auditability, introduce their own tradeoffs. Strict failure on any unresolved element prevents silent downgrade, but it also reduces liveness under partial outages of resolution or status services. Policy *P* can define bounded grace windows for explicitly noncritical fields, yet the default posture remains conservative: no acceptance without verifiable evidence. Finally, the model constrains delegation semantics to explicit, non escalating DG chains. This increases analyzability and safety but excludes informal or heuristic delegation patterns; such patterns can be layered on top using for example policy side conditions but do not alter the core verifier.

## IV. IMPLEMENTATION AND TEST

To test the conceptual framework in a way that is both realistic and manageable, the prototype instantiates a single hybrid deployment profile that concentrates the most relevant technical tensions. Rather than attempting to cover every possible protocol and cryptographic suite, the implementation focuses on a narrow but representative slice where web federated identity (OIDC/VC-JWT) and SSI oriented flows (VC-LD, StatusList2021) coexist under a common verification service and share the same delegation semantics.

Concretely, the prototype adopts what can be described as a Hybrid Minimal Profile:

- On the credential dimension, the system supports both VC-JWT and VC-LD as input formats. VC-JWTs follow the JOSE stack and typical OIDC4VC encodings, whereas VC-LDs are modeled according to W3C VC Data Model 1.1 with BBS+-style linked data proofs and StatusList2021 for revocation. In the current implementation, VC-LDs are fully normalized into the Canonical Verification Context (CVC), but the cryptographic verification of BBS+ signatures is stubbed behind a well defined interface, reflecting library constraints documented in the prototype code.

- On the delegation dimension, Delegation Grants (DGs) are encoded as compact SD-JWT tokens in the web-federated profile and as JSON-LD structures in the SSI profile, but both variants are normalized into the same DG data model.

- On the transport dimension, the prototype exposes a primary/verify endpoint that accepts a normalized VerificationRequest (as JSON) and an auxiliary /verify/oidc4vp endpoint that receives an OIDC4VP vp_token, normalizes it, and then routes it through the same verification pipeline. The full interactive OIDC4VP authorization dance (authorization request, nonce exchange, redirect handling) is out of scope, but the crucial normalization step from VP token to CVC is implemented and exercised in the tests.

- On the anchoring dimension, delegation chains can optionally be bound to a mock blockchain anchor. The prototype implements an in-memory ledger abstraction that stores chain fingerprints, status flags, and expiry times, mirroring the semantics expected from a permissioned blockchain. This keeps the design aligned with the integration of blockchain alternatives thesis scope while avoiding the operational complexity of deploying and maintaining a full DLT network due to the increase in complexity for similar research's [15], and it allows controlled experiments on the latency and overhead introduced by anchoring.

From an evaluation perspective, this hybrid profile also respects the pragmatic constraints defined in the research proposal: a prototype, deployable on a single workstation, with controlled dependencies and reproducible metrics. The verification service is implemented as a FastAPI application exposing a small, stable HTTP surface (/verify, /verify/oidc4vp, /health), backed by a modular verification engine and a metrics framework that records end-to-end latency, normalization cost, verification cost, and invariant satisfaction for each request.

By fixing this Hybrid Minimal Profile as the only deployment configuration, the experiments can concentrate on what matters for the thesis claims:

- whether CVC normalization actually makes the verification pipeline format- and protocol-independent in practice,

- whether delegation chains expressed as DG sequences behave as specified in Chapter 3 under realistic cryptographic and revocation checks, and

- how much additional overhead is introduced when a blockchain-style anchor is required for delegation integrity.

*Prototype Setup*

The prototype instantiates the conceptual architecture as a minimal yet complete end-to-end system that can be executed locally through the following repository: https://github.com/DarSaavedraM/Interoperable-Architecture-for-Digital-Identity-Delegation-for-AI-Agents-with-Blockchain-Integration.git . Its purpose is not to approximate the model, but to realize a concrete, reproducible implementation of the Trust Gateway (L3), the Verification Engine (L1/L2), the Canonical Verification Context (CVC), and the Delegation Grants (DG) data model, under controlled conditions and with explicit limitations.

At runtime, the system is deployed as a Python 3.x service exposing an HTTP API. The Trust Gateway is implemented as a FastAPI application that offers a primary /verify endpoint for direct JSON requests and a secondary /verify/oidc4vp endpoint for normalized OIDC4VP payloads. Both endpoints accept a credential presentation, an optional delegation chain, and a policy identifier, and then construct a CVC that encapsulates all verification-relevant information. The gateway runs in offline mode: all keys, policies, fixtures, status documents and blockchain anchors are loaded from the local file system, ensuring deterministic behavior and strict reproducibility of experiments.

The Verification Engine is implemented as a separate module that receives a CVC from the gateway and executes the deterministic verification function. It performs structural validation of the context, cryptographic verification of VC-JWT credentials using Ed25519 and JWS/JWT primitives, basic normalization of VC-LD credentials to demonstrate format independence, evaluation of DG chains encoded as SD-JWT tokens, status checks against StatusList2021 documents, and policy enforcement. A mock blockchain anchor component is integrated as a thin abstraction that stores delegation chain fingerprints in a local JSON file and validates them during verification whenever the active policy requires anchoring. This design preserves a first interest in providing an architectural role of blockchain for immutable anchoring of delegation chains without coupling the prototype to a specific ledger or node infrastructure.

The experimental environment is completed with a curated set of fixtures and policies. A script initializes all test data: Ed25519 key pairs for issuers, holders, delegates and verifier; JWKS representations of public keys; a small family of VC-JWT credentials and normalized VC-LD examples; DG chains of varying depth and scope; StatusList2021 documents representing active and revoked states; and a set of verification requests that correspond directly to the scenarios defined later in this chapter. Policies are expressed as simple JSON configurations that capture typical deployment choices: issuer allow-lists, required assurance levels, scope and delegation rules, freshness windows for status information, and whether blockchain anchoring is mandatory for a given verification. These components dependencies can be described as in Figure 2.

Finally, the prototype is instrumented with a metrics layer, for every verification request, execution times, result codes, profile classification (federated, SSI, hybrid, anchored), and basic size metrics of inputs and outputs. This instrumentation is passive: it does not alter the verification logic, but simply observes it, enabling the analysis in later sections. Within these boundaries and explicit limitations, like for example, simplified VC-LD verification and a mock anchor store, the setup is sufficient to exercise the core properties of the proposed architecture and to evaluate, under realistic yet controlled conditions, whether verifiable delegation remains deterministic, interoperable, and robust across heterogeneous identity scenarios.

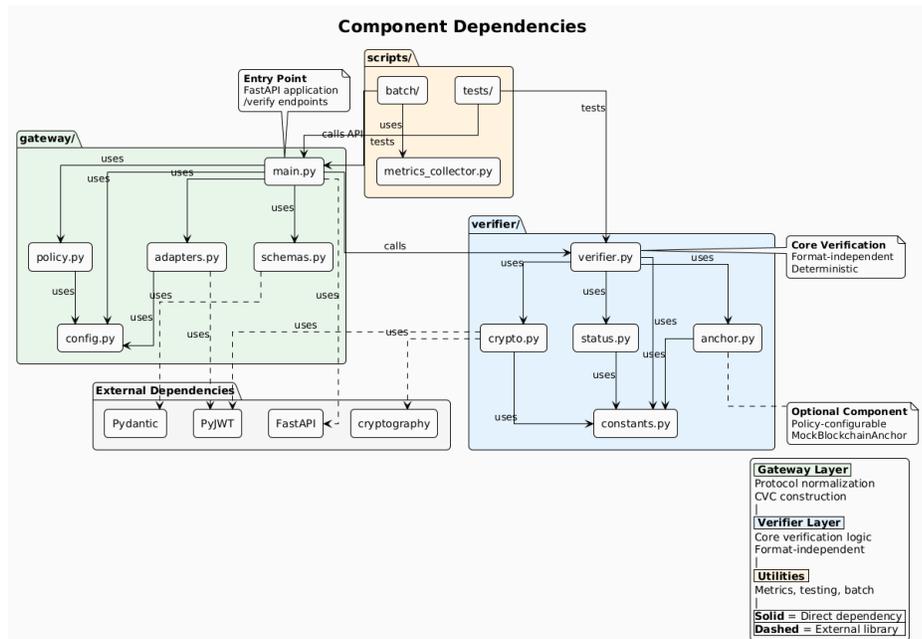

*Figure 3 Components Dependencies*

*Test Scenarios and Procedures*

The implementation exercises five canonical scenarios named S1, S2, S3, S4, S5, that together cover web federated flows, SSI native flows, human to agent delegation, negative/adversarial conditions, and blockchain anchored delegation. Each scenario is implemented as an executable test script that drives the same /verify (and, where relevant, /verify/oidc4vp) endpoint used in normal operation, ensuring that tests measure the real verification pipeline rather than artificial stubs.

    A. Scenario S1: Federated Web Flow (VC-JWT / OIDC4VP)

S1 represents the "least disruptive" adoption path for existing web and enterprise identity systems. The verifier receives a VC-JWT issued under a JOSE/JWK infrastructure, optionally wrapped in an OIDC4VP response. The goal is to show that such inputs are normalized into a Canonical Verification Context and verified without any SSI specific assumptions.

The test script test_s1_federated.py constructs requests where the presentation field contains a VC-JWT, together with a policy identifier and, in some variants, a short delegation chain expressed as SD-JWT DGs. The Trust Gateway detects the VC-JWT profile, verifies the JWS signature with the issuer's JWKS, checks temporal bounds and StatusList2021 status, and builds a CVC that is then passed to the verification engine. The test asserts that:

• Verification succeeds (OK result code) when signatures, status, and policy are consistent.

• Any manipulation of the JWT payload or header yields an E200 proof-error response.

• For OIDC4VP inputs, the extraction of the vp_token and subsequent normalization leads to the same CVC as the direct VC-JWT case, confirming protocol independence for the same underlying credential.

    B. Scenario S2: SSI Native Flow (VC-LD, StatusList2021)

S2 exercises the SSI centric side of architecture. Here, the verifier receives a VC expressed in JSON-LD, accompanied by a StatusList2021 URL and, depending on the variant, either a DG encoded as LD-Proof or an SD-JWT DG used as a bridge between ecosystems.

The script test_s2_ssi.py reads VC-LD fixtures and constructs requests that mimic a typical wallet-to-verifier presentation. In the prototype, the LD proof is treated as an opaque, structurally validated object; full BBS+ verification is documented as a limitation and delegated to external SSI libraries in future work. The Gateway normalizes the VC-LD into a CVC, resolving issuers and subjects, importing the status reference, and preserving the proof type so that the verification engine can distinguish LD flows from JOSE ones.

The test asserts three core properties: that normalization of VC-LD produces a CVC structurally equivalent to the VC-JWT case for the same claims; that StatusList2021 documents are fetched and parsed correctly; and that revocation flags lead to fail-closed behavior when a credential is marked as revoked in the status list.

    C. Scenario S3: Hybrid Human and Agent Delegation

S3 focuses on the central first intuitive research question: can a human principal delegate authority to an AI agent across heterogeneous identity stacks in a verifiable, bounded, and auditable way?

The script test_s3_hybrid.py exercises chains where a human holder receives an initial VC-JWT, then issues one or more Delegation Grants (DGs) to an AI agent. DGs are encoded either as SD-JWTs or as LD objects, depending on the variant, and include explicit scope restrictions and temporal bounds.

The test sends verification requests where the acting subject is the agent, and the delegation chain $\Delta$ is provided alongside the credential. The verification engine reconstructs the delegation chain, validates each DG signature, checks that each step narrows or preserves scope, and enforces that the chain is acyclic and anchored in a principal authorized by the root credential.

    D. Scenario S4: Negative and Adversarial Cases

S4 is designed to falsify the model by construction. Instead of expecting success, each test case encodes a distinct violation of the rules defined in Chapter 3.

The script test_s4_negative.py covers four principal categories of failure:

• Structural errors, such as malformed CVC fields or missing mandatory attributes, which must return FormatError (100).
• Cryptographic errors, such as tampered signatures or unknown issuers, which must return E200.
• Delegation-specific errors, including scope escalation and broken chains, which must trigger E300 or E400 depending on whether the issue is structural or semantic.
• Policy violations, such as untrusted issuers or disallowed credential types, which must generate E500.

Each negative fixture is derived from a valid base case by applying a minimal mutation, flipping a bit in the signature, changing a scope value, forcing a DG expiry, or corrupting the chain order. The test asserts that the verifier returns the expected error code and that no invalid request is ever accepted as OK. This scenario is essential to substantiate the fail-closed design claim and to demonstrate that adding new profiles or formats does not silently widen the accepted attack surface.

The test suite confirms that valid anchored chains are accepted, that unanchored chains are rejected when anchoring is required, and that any inconsistency between the computed fingerprint and the stored anchor results in an E300 chain error response. This scenario demonstrates how blockchain integration strengthens accountability without changing the core CVC and DG semantics.

*Batch Execution Procedure*

Finally, the batch procedure validates that the verification

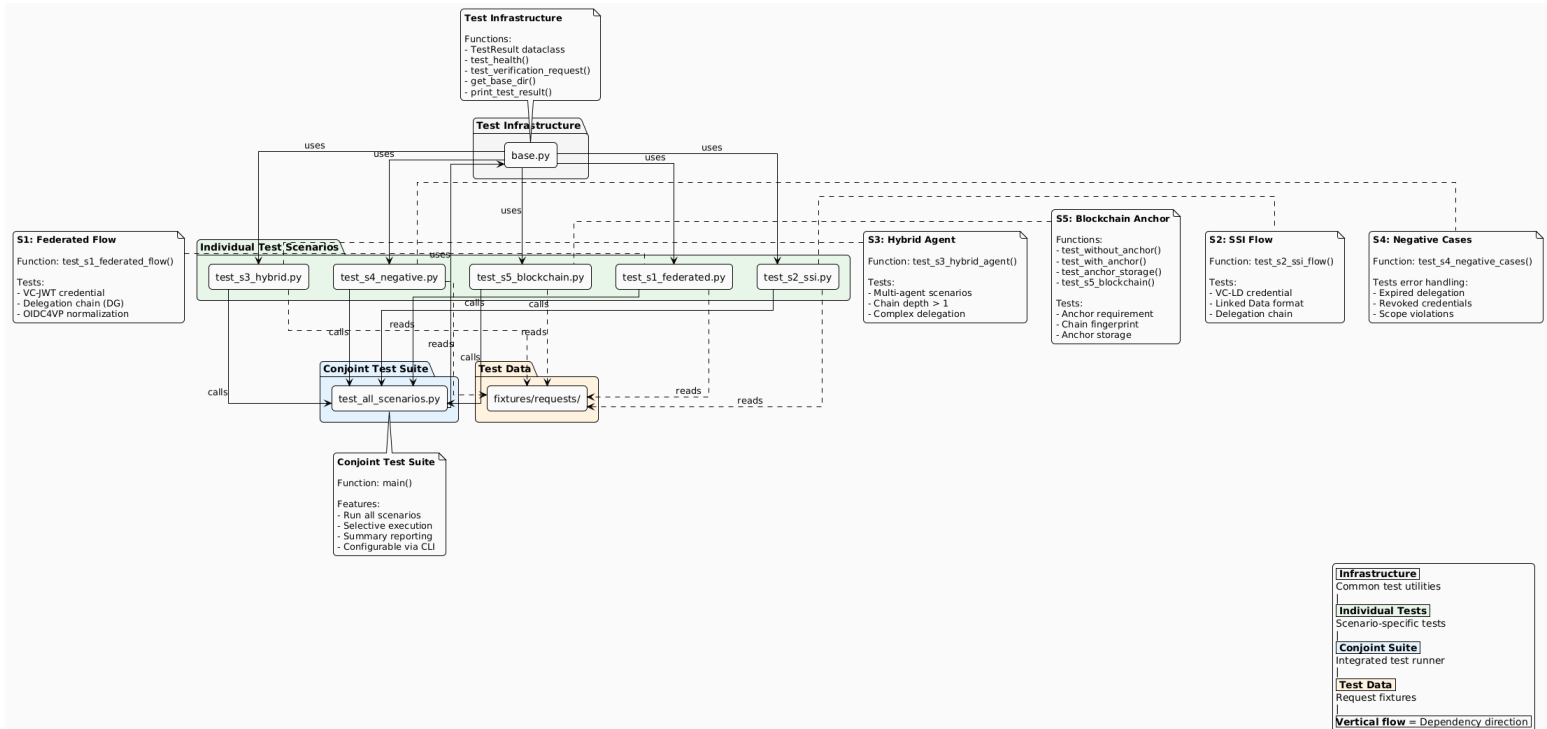

*Figure 4 Test scenarios separation*

E. Scenario S5: Blockchain Anchored Delegation

S5 exercises the blockchain related component of the architecture. The objective is not to benchmark a specific ledger, but to validate that delegation chains can be anchored to an immutable log and that verification logic reacts correctly when anchors are present, missing, or inconsistent.

The script test_s5_blockchain.py uses the mock blockchain implementation provided by verifier/anchor.py, which stores chain fingerprints, timestamps, and synthetic "block identifiers" in a JSON file under fixtures/anchors/. Two classes of tests are executed. First, verification runs with require_anchor=false, ensuring that normal delegation works even without anchoring. Then, verification is repeated with require_anchor=true, which forces the verifier to compute the fingerprint of the delegation chain, compare it with the stored anchor, and fail if no matching anchoring record exists or if the anchor is marked as revoked or expired.

pipeline behaves consistently under load and across a broad distribution of inputs. Using generate_batch.py, the system creates 1 200 synthetic requests that randomly mix S1 to S3 patterns with varying delegation depths and with or without blockchain anchoring. The script run_batch_metrics.py replays this corpus against a running server, collecting a metrics object for each verification and aggregating the results into JSON and Markdown reports. It also serves as an additional robustness check: all 1 200 requests must satisfy the invariants encoded in the verification engine, and any deviation would immediately surface as a failed batch run.

*Metrics and Empirical Results*

The batch experiment comprised 1200 verification requests executed against the prototype, covering both supported credential profiles (VC-JWT and VC-LD/BBS+), delegation chains of depth 0 to 3, and optional blockchain anchoring for a subset of delegated flows. All metrics reported in this section are derived from the actual execution traces produced by the

/verify endpoint and analyzed by the metrics scripts described earlier in this chapter.

At a global level, the system achieved a 100 % success rate with no failed verifications, an average end-to-end (E2E) latency of 58.60 ms (median 34.47 ms, P95 196.72 ms), and an average normalization time of 2.68 ms and verification time of 28.98 ms per request. This confirms that the prototype can process heterogeneous inputs deterministically and within latency ranges compatible with interactive applications.

*Latency and Scalability by Delegation Depth*

Verification latency increases approximately linearly with depth, while normalization remains almost constant and low.

Across depths 0 to 3, normalization time remains close to 2-3 ms, while verification cost grows from roughly 7 ms at depth 0 to about 51 ms at depth 3. The observed latency increase is well approximated by a linear coefficient of ≈ 14 ms per additional depth level (with low variance), confirming the expected behavior of the delegation chain algorithm.

This behavior is consistent with the conceptual design in Chapter 3: each additional link in the delegation chain introduces one extra signature verification, scope check, temporal validity check, and (optionally) blockchain-anchor verification, but no global recomputation or backtracking across the chain. The fact that depth 3 remains under 80 ms E2E on average suggests that, for moderate chain lengths, the approach might be compatible with real-time verification requirements in typical web and mobile scenarios, but uncertain for more complex scenarios.

| Depth | Requests | E2E mean (ms) | E2E median (ms) | E2E P95 (ms) | Normalization mean (ms) | Verification mean (ms) |
| --- | --- | --- | --- | --- | --- | --- |
| 0 | 286 | 36.60 | 21.71 | 131.35 | 2.28 | 7.15 |
| 1 | 323 | 47.09 | 33.58 | 165.06 | 2.52 | 17.53 |
| 2 | 312 | 73.03 | 44.18 | 242.78 | 2.79 | 40.89 |
| 3 | 279 | 78.37 | 44.18 | 242.78 | 2.90 | 51.28 |

**Table 1 Latency by Delegation Depth (1 200 requests)**

*Size Overhead and Canonical Expansion*

The architecture deliberately separates compact credential formats at the edge (VC-JWT / VC-LD) from a richer internal representation, the Canonical Verification Context (CVC). The metrics allow us to quantify the overhead introduced by this normalization step.

**Table 2 Size Metrics for Key Artifacts**

| Artifact | Samples (non-zero) | Mean size (bytes) | Median size (bytes) | Notes |
| --- | --- | --- | --- | --- |
| VC-JWT | 565 / 1 200 | 644 | 638 | Compact edge credential |
| CVC (serialized) | 1 200 / 1 200 | 5 442 | ≈ 4 400 | Normalized internal representation |
| VRO-JWT (result) | 1 200 / 1 200 | 525 | 543 | Signed verification result token |
| DG chain (if any) | 914 / 1 200 | 2 837 | 2 904 | Serialized delegation grants in request |

On average, the CVC is about 8.4× larger than the original VC-JWT, reflecting the inclusion of structured metadata, explicit status references, and a normalized representation of the delegation chain. This overhead is consistent with the design goal: the CVC is not meant for only transmission across domains, but as a rich, internal, envelope that allows the verifier to apply uniform logic regardless of the originating profile or proof format.

At the same time, the verification result object (VRO-JWT) remains compact (~0.5 KB on average), which is important for downstream systems that need to store audit trails or cache verification decisions without incurring significant storage costs. The DG-chain size, around 2.8 KB on average, is comparable to or smaller than typical JWT access tokens used in web systems, indicating that explicit, structured delegation can be encoded without prohibitive overhead for moderate chain lengths.

*Correctness Invariants and Determinism*

Beyond raw performance, the experiment was designed to test whether the prototype faithfully enforces the logical invariants specified in Chapter 3. The metrics framework records, for each request, whether core invariants are satisfied: scope containment, temporal validity, signature correctness, chain integrity, and structural validity.

**Table 3 Invariant Pass Rates (1 200 requests)**

| Invariant | Passed / Total | Pass rate |
| --- | --- | --- |
| Scope containment | 1 200 / 1 200 | 100 % |
| Temporal validity | 1 200 / 1 200 | 100 % |
| Signature verification | 1 200 / 1 200 | 100 % |
| Delegation-chain integrity | 1 200 / 1 200 | 100 % |
| Structural validity (schema) | 1 200 / 1 200 | 100 % |

In this batch, all invariants passed for all requests, including those with non-trivial delegation chains. This indicates that, under the tested conditions, the implementation is consistent with the formal model: no credential is accepted without valid signatures and temporal bounds; no delegation chain is accepted if it breaks scope containment or chain-integrity constraints; and no malformed request bypasses structural validation.

Determinism was evaluated through the uniqueness and reproducibility of verification-result hashes. All 1200 VRO hashes were unique, and repeated executions with identical inputs produced identical outputs (once timestamps and nonces were fixed). This behavior is aligned with the requirement that the verification function behaves as a pure function of the Canonical Verification Context: given the same credentials, delegation chain, status state, and policy, the same result is produced.

*Blockchain Anchor Impact*

Finally, the experiment quantified the impact of enabling blockchain anchoring for delegation chains. Approximately half of the requests with delegation chains required a blockchain anchor check, while the other half relied solely on off-chain status mechanisms.

**Table 4 Latency with and without Blockchain Anchor**

| Mode | Requests | E2E mean (ms) | E2E median (ms) | E2E P95 (ms) | Verification mean (ms) |
|---|---|---|---|---|---|
| With anchor | 455 | 77.79 | 44.18 | 242.78 | 49.74 |
| Without anchor | 459 | 53.30 | 33.58 | 165.06 | 21.99 |
| Anchor overhead | ------- | +24.49 | +10.60 | +77.72 | +27.75 |

On average, enabling blockchain anchors increases E2E latency by about 24.5 ms (+45.9 %) and more than doubles the verification component (+126 %). The overhead grows with chain depth, from roughly +12.5 % at depth 1 to over +60 % at depth 3. This is consistent with the fact that each delegated link that requires anchoring triggers at least one additional on-chain read or corresponding cache lookup.

From a design perspective, these results show that blockchain anchoring is viable but not "free": it provides stronger tamper-evidence for delegation chains at a measurable cost in latency. The architecture's choice to make anchoring an optional, policy-driven feature rather than a hard requirement is therefore empirically justified. Systems that prioritize latency over maximal auditability can disable anchoring or apply it selectively, while high assurance environments can tolerate the additional overhead in exchange for stronger guarantees.

*Summary of Empirical Findings*

Taken together, the metrics support three core claims of the proposed model:

1. Format-agnostic verification is practical: VC-JWT and VC-LD/BBS+ inputs are normalized to a common CVC with negligible differences in normalization time ($\approx$ 4.2 % variation), and the verification logic is truly independent of the original format or transport.

2. Delegation chain validation scales linearly with depth under the implemented model, remaining within interactive latency bounds for chains up to depth 3, while enforcing all structural, cryptographic, temporal, and scope invariants.

3. Blockchain anchoring provides an effective, optional strengthening of integrity guarantees at a moderate but measurable latency cost, which can be tuned via policy according to the needs of each deployment.

These findings do not exhaust all possible evaluation scenarios, but they provide a concrete, reproducible baseline demonstrating that the conceptual framework can be instantiated in code and subjected to empirical scrutiny under realistic conditions.

*Discussion of Findings and Prototype Limitations*

The prototype provides concrete evidence that the conceptual framework is implementable as an end-to-end verification service and that its main design choices behave as intended under realistic, though controlled, conditions. The five scenario families and the 1200 request batch collectively exercised all key elements of the model: Canonical Verification Context (CVC) normalization, protocol-agnostic verification, explicit Delegation Grants (DGs), and optional blockchain anchoring. Across all runs, the verifier maintained a 100 % success rate in enforcing its invariants (where "success" here means correct acceptance or rejection according to the scenario, not universal approval of requests), with no observed violations of structural, cryptographic, temporal, or delegation constraints.

End-to-end latency remains in a range compatible with interactive use (tens of milliseconds on average, with the 95th percentile below 200 ms in the tested environment), and normalization accounts for only a small, stable fraction of this cost. The dominant factor is the verification phase, which includes signature checks, status resolution, and delegation-chain evaluation. Crucially, latency grows approximately linearly with delegation depth. This matches the algorithmic structure: each additional DG introduces one more local verification step but does not trigger any global recomputation or search over the chain. Within the tested depths (0 to 3), this linear growth keeps total latency within reasonable bounds, supporting the thesis that explicit, structured delegation can be made operational without prohibitive performance penalties for realistic chain lengths.

Blockchain anchoring behaves as a controlled, policy-driven strengthening of integrity guarantees. When anchoring is disabled, the system operates as a conventional off-chain verifier, relying on standard status mechanisms. When anchoring is required, the verifier computes a fingerprint of the delegation chain, consults the anchor store, and rejects any chain that is unanchored, revoked, or inconsistent with the recorded hash. The measured overhead on the order of a few tens of milliseconds on average, increasing with chain depth is significant but not destabilizing. It shows that ledger-backed integrity can be integrated as an optional layer rather than a hard requirement, leaving deployers free to trade latency for auditability according to their context. This reinforces one of the central design choices of the architecture: blockchain is used as an anchoring and accountability mechanism, not as the universal substrate for all identity operations.

At the same time, the prototype exposes clear limitations that constrain how these findings should be interpreted. Some limitations are cryptographic. VC-LD credentials are fully normalized and treated as first-class inputs, but BBS+ signature verification is abstracted behind an interface rather than implemented with production libraries in this prototype. Holder binding and proof-of-possession are represented structurally (via key identifiers and JWKS entries) rather than via interactive DPoP or mutual TLS exchanges. StatusList2021 handling follows the intended semantics but uses a simplified decoding path instead of a fully optimized bitstring implementation. These choices do not invalidate the architectural claims since the interfaces and invariants are present and exercised but they mean that the current prototype should be regarded as a proof of concept rather than a complete SSI verifier.

Other limitations concern the evaluation setting. All experiments were run in a single node environment, under a fixed hardware and software stack, with synthetic but controlled request distributions and without adversarial load, network partitions, or multi tenant contention. The batch generator explores a range of profiles and delegation depths, yet real deployments would face far greater heterogeneity in issuer practices, schema evolution, wallet behavior, and policy regimes. Likewise, while the prototype enforces a clear fail posture and exposes a disciplined error taxonomy, it has not undergone formal verification, fuzzing, or third party security review. The current results demonstrate that the *model* is implementable and behaves coherently under the tested conditions; they do not constitute a claim that this particular implementation is ready for deployment in high assurance environments.

Taken together, empirical evidence does what is required for the purposes of this thesis. It shows that the proposed architecture for verifiable delegation, built around a Canonical Verification Context, explicit Delegation Grants, and an optional blockchain anchoring layer can be realized in running code, that it behaves deterministically across heterogeneous identity formats, and that its main trade-offs (delegation depth, anchoring, normalization overhead) are measurable and controllable. At the same time, the explicit limitations of the prototype delineate the boundary between the validated conceptual contribution and the engineering work that remains for large-scale, production deployments. The next chapter builds on this foundation to discuss the broader implications of the model, its position within the digital identity landscape, and its potential extensions.

## V. CONCLUSIONS

The resulting evidence supports three central claims. First, format and protocol independence is practically achievable: once inputs are reduced to CVCs, the verifier processes VC-JWT, VC-LD, OIDC4VP presentations, and SD-JWT/LD-based DGs through the same logic, without specific branches for profiles. Second, delegation chains expressed as explicit DG sequences remain tractable and predictable: verification cost grows approximately linearly with depth, all structural, cryptographic, temporal, and scope containment invariants hold for the tested chains, and latency stays within interactive ranges for realistic depths. Third, blockchain anchoring can be integrated as an optional integrity layer that strengthens accountability for delegation chains at a measurable but controlled latency cost, without changing the semantics of DGs or the structure of the CVC.

At the same time, the prototype clarifies the scope of validation. Only a subset of DID methods, proof suites, and revocation mechanisms is implemented; some cryptographic operations (such as BBS+ verification and interactive holder binding) are abstracted behind interfaces rather than fully realized; and all experiments are conducted in a single node, offline, non adversarial environment. The results therefore show that the proposed model is internally consistent and implementable, and that its main trade offs can be measured and tuned, but they do not claim completeness across all standards or deployment contexts.

*Theoretical Contributions*

The first contribution is the use of the CVC as a normalization abstraction that makes verification a well-defined function over a single, protocol agnostic object. This goes beyond the first impression notion of supporting multiple profiles: it states that verification itself can be specified and reasoned independently of both encoding and protocol.

Closely related is the explicit format and protocol independence of verification as properties, not just as implementation conveniences. By forcing all inputs through the CVC, the thesis shows that the same verification pipeline can handle heterogeneous ecosystems (W3C VC-JWT, VC-LD/BBS+, SD-JWT) without duplicating logic, and that performance differences between formats are confined to the normalization phase. This is a theoretical contribution in the sense that it reframes interoperability: instead of asking "how do we map one profile to another?", it asks "what is the minimal common verification state such that profiles become mere encodings of that state?".

A second contribution concerns delegation as a first class, bounded construct, rather than an emergent property of tokens or sessions. Delegation Grants, are not just a new type of credential; they are a way to separate descriptive statements (VCs) from normative ones (who may act on whose behalf). The theoretical contribution here is twofold. On one hand, DGs embody a precise notion of authority reduction along a chain: each step must be no more permissive than its predecessor, which gives a clear semantics to "safe" delegation and rules out privilege escalation by construction. On the other hand, DGs provide a uniform way to talk about delegation whether the acting entity is a human or an AI agent. The Delegation Tetrahedron generalizes the classic trust triangle exactly to support this: it makes the path from issuer and principal to agent and verifier explicit, rather than leaving it implicit in bearer tokens or client implementations.

A third contribution is the treatment of blockchain as an optional integrity layer rather than as a foundational substrate, which was the first intuition of the research, this decision was made after reviewing the underlying technologies available that ultimately can be articulated without a ground approach on Blockchain. The theoretical value of this design is to separate two concerns that are often conflated: cryptographic validity of credentials and delegations on the one hand, and immutable recording of their state on the other. By keeping DG semantics and CVC structure independent of any specific ledger, the model can absorb or discard blockchain anchoring without changing its core logic.

Finally, the research treats deterministic, reproducible verification as a design objective rather than an incidental property of a particular implementation. Once verification is defined as a pure function over the CVC, it becomes possible to require that identical inputs always yield identical outputs, and to use signed Verification Result Objects and chain fingerprints as stable artefacts for logging and audit. This is not a new cryptographic idea, but in the context of digital identity it formalizes an often implicit requirement: that verification outcomes can be reproduced, compared, and audited across time and implementations.

These elements, taken together, provide a coherent lens for specifying and analyzing verifiable delegation in identity systems that span both human users and autonomous agents adding value to the development of a solid infrastructure for artificial intelligence management and responsibility for the users.

*Practical Implications*

The model developed in this thesis has direct consequences for how identity systems can be designed, extended, and integrated in practice. The key point is that the CVC, DGs, and the layered architecture do not stay at the level of abstraction: they map cleanly to implementation choices that existing ecosystems can adopt without discarding their current investments.

A first implication is for interoperability between federated and SSI ecosystems. In the current landscape, deployments tend to choose either an OIDC-centric stack (VC-JWT, SD-JWT, OIDC4VCI/VP) or an SSI-centric stack (VC-LD, BBS+, StatusList2021, DIDComm), and interoperability is treated as a profile mapping problem. The CVC based approach suggests a different path: verifiers can be built around a single internal verification state, with protocol and format differences confined to the Trust Gateway and its adapters. In practical terms, this means that an institution can accept, for example, both a bank issued VC-JWT via OIDC4VP and a wallet-presented VC-LD from an SSI network, and evaluate them using the same verification engine and policy layer. The cost of supporting an additional profile then becomes the cost of writing one more adapter, not of replicating the entire verification stack.

A second implication concerns delegation in systems that incorporate autonomous agents. Today, many architectures treat agents, whether they are bots, RPA components, or LLM-based service, as opaque clients authenticated via API keys or long lived tokens. The DG model provides a concrete alternative: agents receive explicit, verifiable delegations that encode who authorized them, for which actions or resources, and for how long, with the guarantee that those delegations cannot silently expand their privileges beyond what was granted.

A third implication is the flexible position of blockchain in identity deployments. The architecture shows that ledgers can be reserved for anchoring and audit, while the core verification logic remains fully functional off-chain. For practitioners, this translates into deployable options rather than a binary choice "with blockchain" or "without blockchain". Environments that require immutable evidence of delegation decisions such as cross jurisdiction regulatory contexts or multi-party consortia can enable anchoring for specific classes of delegations or policies, accepting the additional latency cost in exchange for stronger guarantees. Other environments can omit anchoring entirely, or restrict it to exceptional flows, and still benefit from the same CVC and DG machinery.

Finally, the insistence on deterministic, reproducible verification has operational implications for testing, auditing, and governance. Because the verification function is defined over the CVC and is implemented in a way that produces stable outputs given the same inputs, verification results can be logged, hashed, and compared across time and implementations. This simplifies regression testing when upgrading libraries or protocols, supports independent re-verification of past decisions, and provides a concrete basis for audit trails in which both credentials and delegation chains can be re-evaluated under the same or updated policies. For system operators, this means that verification can be treated as a controlled, inspectable process rather than as a black box buried inside protocol handlers.

In sum, the practical message of the research is that verifiable delegation does not require discarding existing standards or committing to a single ecosystem. Instead, it suggests that identity infrastructures can be evolved by adding a normalization layer (CVC), an explicit delegation primitive (DG), and a clear separation between verification logic and transport protocols. These elements provide concrete hooks for introducing AI agents safely, bridging SSI and federated identities, and incrementally adopting blockchain integrity where it is justified, without redefining the entire stack.

*Future work*

Due to the limitations of the prototype, it is necessary to enhance the implementation of all the proposed models as well as the deployment on real world scenarios to prove the consistency of the information after a real data connection. In this way there's 3 different areas of further research identified:

a. Integration and testing of the architecture with real data.
b. Deployment of delegation chains in main net of large blockchains.
c. Development of integration services for existing AI Agents.

This approach will resolve some of the limitations of the approach in the research. The first proposed line is to prove the correctness of the information using real profile data and existing technologies such as has been made on the prototype. This is a necessary step before the production of a worldwide accepted protocol and will also provide information about necessary adjustments.

In the other hand, to include information about cost, time and gas fees for a further integration of blockchain solutions it is necessary to include the development of smart contracts and its deployment. In this way it may be easier to measure the impact of such integration and provide more information for developers of the decentralized sphere.

Finally, the goal of the technical development must be an easy and sharp experience for the final user. In this sense, it's necessary to provide an easy to integrate solution for AI Agents providers in a way that the user doesn't get unnecessary friction with the services while adding value to companies, institutions and users.

Figure List

Table List